\documentclass[conference]{IEEEtran}
\IEEEoverridecommandlockouts
\usepackage{cite}
\usepackage{amsmath,amssymb,amsfonts}
\usepackage{algorithmic}
\usepackage{graphicx}
\usepackage{textcomp}
\usepackage{xcolor}
\usepackage{float} 
\usepackage{subfigure}
\usepackage{amssymb}
\usepackage{booktabs}
\usepackage{makecell}
\usepackage{multirow}
\usepackage{cite}
\usepackage{algorithm} 
\usepackage{amsthm}
\usepackage{array}
\usepackage{color}

\def\BibTeX{{\rm B\kern-.05em{\sc i\kern-.025em b}\kern-.08em
    T\kern-.1667em\lower.7ex\hbox{E}\kern-.125emX}}
\begin{document}


\title{GPLA-12: An Acoustic Signal Dataset of Gas Pipeline Leakage\\

}

\author{\IEEEauthorblockN{ Jie Li}
\IEEEauthorblockA{\textit{School of Intelligent Technology and Engineering} \\
\textit{Chongqing University of Science and Technology}\\
Chongqing, China \\
jieli@cqust.edu.cn}


\and
\IEEEauthorblockN{ Lizhong Yao}
\IEEEauthorblockA{\textit{School of Electrical Engineering} \\
\textit{Chongqing University of Science and Technology}\\
Chongqing, China \\
yaolizhong225@163.com}
}

\maketitle

\begin{abstract}
In this paper, we introduce a new acoustic leakage dataset of gas pipelines, called as GPLA-12, which has 12 categories over 684 training/testing acoustic signals. Unlike massive image and voice datasets, there have relatively few acoustic signal datasets, especially for engineering fault detection. In order to enhance the development of fault diagnosis, we collect acoustic leakage signals on the basis of an intact gas pipe system with external artificial leakages, and then preprocess the collected data with structured tailoring which are turned into GPLA-12. GPLA-12 dedicates to serve as a feature learning dataset for time-series tasks and classifications. To further understand the dataset, we train both shadow and deep learning algorithms to observe the performance. The dataset as well as the pretrained models have been released at both www.daip.club and github.com/Deep-AI-Application-DAIP.
\end{abstract}


\section{Introduction}
Fault detection and diagnostics, denoted as FDD, is a critical technique to guaratee safety that widely used in industrial processes (e.g., machine manufacturing, chemical operation, pipeline transportation, and software systems) \cite{gao2015survey, nandi2005condition, duan2018deep, amiruddin2020neural}. A fault symptom as a variable is detected through developing inherent relationships between fault and operation parameters. FDD techniques can be divided into two categories: simple-model and complex-model approaches \cite{liu2018artificial, jia2018neural, he2017deep, boussif2020intermittent}. For the former, traditional approaches which discover anomalies of a single parameter, such as overhigh/overlow values and deviation limitation of change rates . For the latter, coopertive influences on fault among multiple parameters are considered and analyzed to find the instrinsic mapping from collected data of given parameters to fault types.

Among these parameters, acoustic signal is a significant tool to observe internal patterns of faults as a non-destructive testing technology \cite{stowell2015detection, kim2009detection, li2019gear, glowacz2018early, glowacz2018acoustic, glowacz2019fault, parey2019gearbox}. Acoustic-based FDD observes fault features with acoustic signals gathered by microphones, mainly used in mechanical faults and pipe leakages. Thereinto, acoustic datasets are the key part of solutions which determine the quality of fault identification. However, most researches adopt private datasets in cooresponding papers and there are only few public acoustic datasets for FDD. 

The aim of this paper is to provide supports of open researches in the field of FDD: 

\begin{enumerate}[]
\item Provide a publicly available dataset of FDD and present the estimations of classification accuacy of this dataset.
 
\item Introduce the data collection system in detail (e.g., hardware parameters) and explain data properties of this dataset.

\item Compare the performance of both simple and complex classifiers on the dataset.

\end{enumerate} 

\section{Background}
ESC-50 and Gearbox Fault Diagnosis Data are popular datasets on classification and FDD as the same purpose of our dataset \cite{zhang2018deep, sailor2017unsupervised, nanni2020ensemble, xie2020zero}.

\textbf{ESC-50 \cite{piczak2015dataset}}: The dataset is labelled with 50 categories (e.g., dog, rain, and door knock) of 2000 audios, which is widely used in environmental sound classification. Each class has 40 examples which contains 5 seconds. The dataset can be divided into 5 major categories: animals, natural soundscapes/water sounds, human/non-speech sounds, interior/domestic sounds, and exterior/urban noises. The top accurary publiced for ESC-50 is 94.1$\%$ by an integration with CNN, RBM and GTSC. Website is $https://github.com/karolpiczak/ESC-50$.

\textbf{Gearbox Fault Diagnosis Data}: The dataset referred to vibration is collected with SpectraQuest's Gearbox Fault Diagnostics Simulator, and publiced in 2018. The vibration information is obtained by 4 vibration sensors installed in 4 directions under 2 scenarios (i.e., healthy condition and broken tooth condition). The load on the simulator covers from 0 to 90 percent. Website is $https://openei.org/datasets/dataset/gearbox-fault-diagnosis-data$.

\section{GPLA-12 Dataset}
\subsection{Data Collection}
Fig. 1 shows a real gas pipeline system with detailed components, which is used to collect leakage acoustic datasets (i.e., GPLA-12). The continuous and stable pipeline gas can be generated through an air compressor unit as described in Fig. 1 (a). Specifically, the displacement of air compressors is 3.5 m$^3$/min and the exhaust pressure of that is 0.85 Mpa. The volume of air tank is 1 m$^3$/min. With an air tank, the compressed air has a buffering place, and can be maintatined at a constant pressure in the pipeline system. In regards to refrigeration dryer in the system, it can reduce water contents around compressed air by lowering gas temperature as working pressure is 0.6$\sim$1.0 Mpa. In Fig. 1 (a), a heatless regenerative adsorption dryer (HRAD) follows the most advanced pressure swing adsorption principles (PSA) in the world, and uses activated alumina and molecular sieve as adsorbents. The HRAD has a two-tower structure, where one tower is for adsorption and the other is for regeneration. The HRAD performs alternative cycles to ensure a continuous drying process of compressed air. For both HRAD and refrigeration dryer, the displacement of air compressor are 3.8 m$^3$/min while working simultaneously. A three-stage filtration treatment in Fig. 1 (a) is used to remove the moisture, dust, and oil in the gas system as much as possible.

\begin{figure}[htp]
\centering
\includegraphics[width=0.49\textwidth, height=0.45\textwidth]{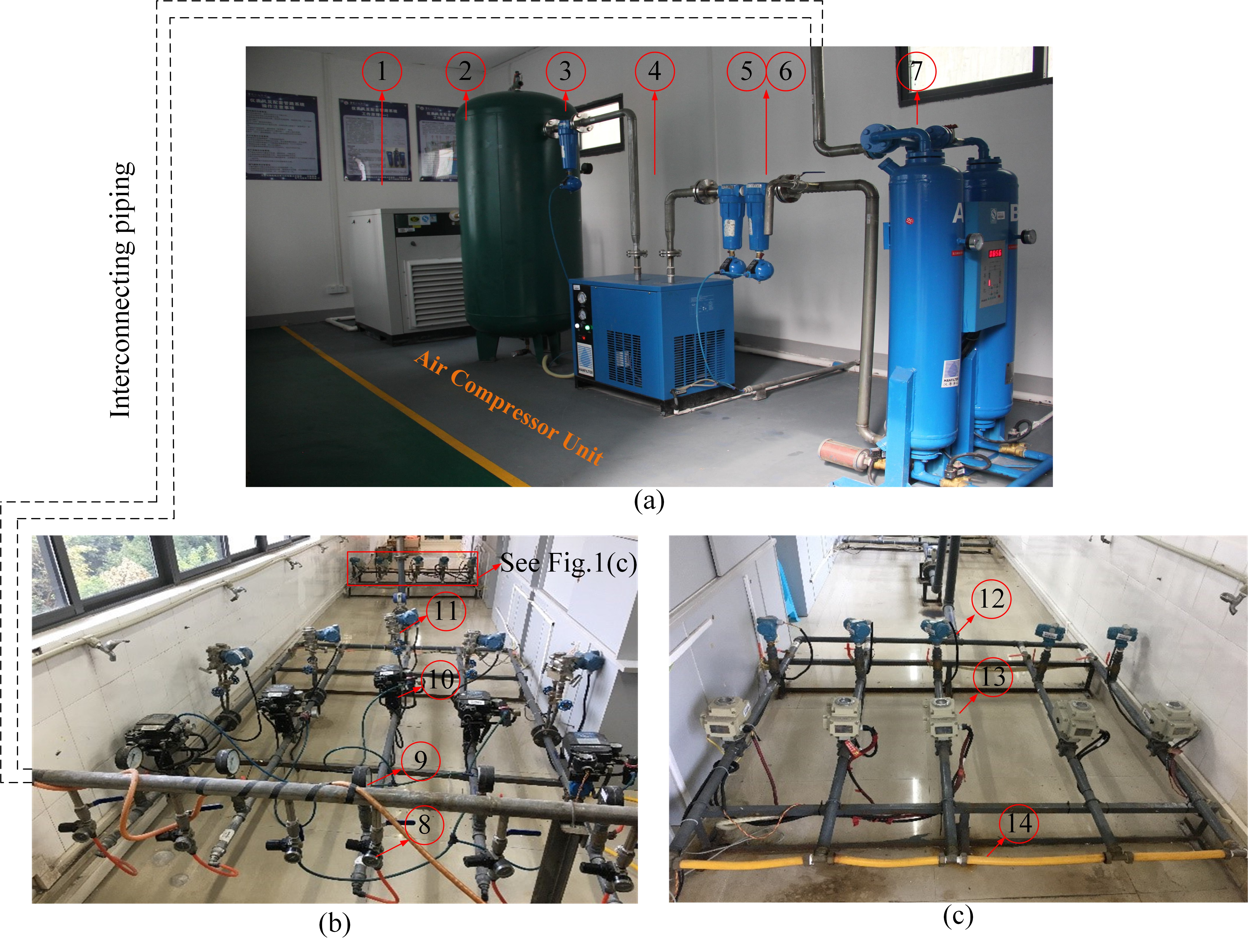}
\caption{Images of the gas pipeline system. (a) Air compressor units; (b) The front end of gas pipeline system; (c) The back end of gas pipeline system. Note that key components in the system are denoted with cooresponding numbers (i.e., 1, screw compressor; 2, air tank; 3, water elimination filter; 4, refrigeration dryer; 5, dust removal filter; 6, oil removal filter; 7, heatless regenerative adsorption dryer (HRAD); 8, pressure regulating valve; 9, pressure gauge; 10, pneumatic control valve; 11, different pressure orifice flowmeter; 12, pressure transmitter; 13, electric control valve; 14, gas discharge tube).}
\end{figure}

\begin{figure}[htp]
\centering
\includegraphics[width=0.46\textwidth, height=0.2\textwidth]{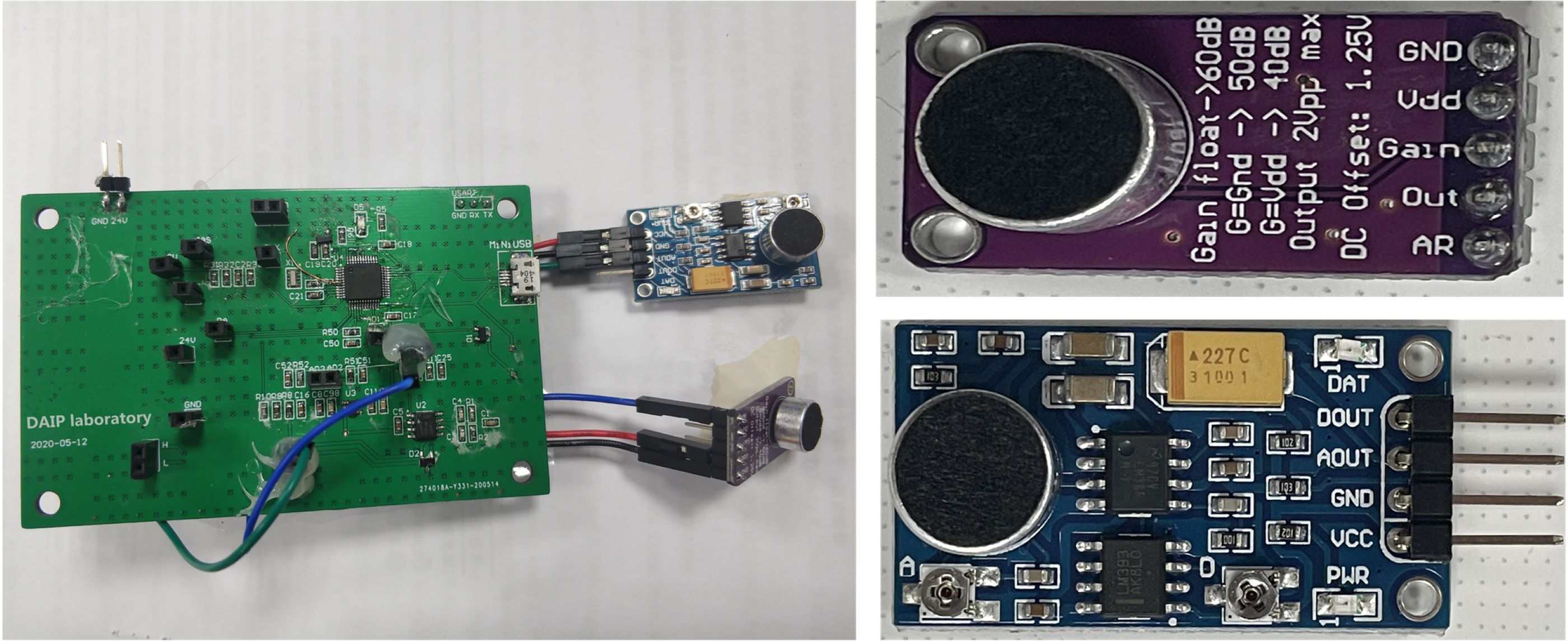}
\caption{The acoustic data collection device.}
\end{figure}

A acoustic signal collection device contains a self-designed signal transmission module, a single chip microcomputer (STM32f103), and two kinds of microphones (illustrated as in Fig. 2). These two microphones have different parameters listed in Table I where the purple one is called as the 1st microphone and the blue one is named as the 2nd microphone. The diversity of microphones leads to the variety of acoustic signals which is dedicated to reducing devices bias. Each acoustic sample sustains 5 seconds and is preprocessed as a discrete and time-series variable with 1460 length.

\begin{table}[htp]
  \caption{Parameters of Microphones.}
  \label{Tab:bookRWCal}  \centering
\begin{tabular}{p{0.5cm}p{1.2cm}p{2.2cm}p{1cm}p{1.4cm}}
  \toprule
   Name  & Version & Voltage  & Sensitivity  &  Signal Range \\
\midrule
 1st  & MAX4466  & $+2.4V\sim+5.5V$  & 112db & $0HZ\sim600kHZ$ \\
     \midrule
2nd & LM386 &  $+4.5V\sim+5.5V$ & 50db & $100HZ\sim10kHZ$\\
    \bottomrule
  \end{tabular}
\end{table}

\begin{figure*}[htp]
\centering
\includegraphics[width=0.88\textwidth, height=0.55\textwidth]{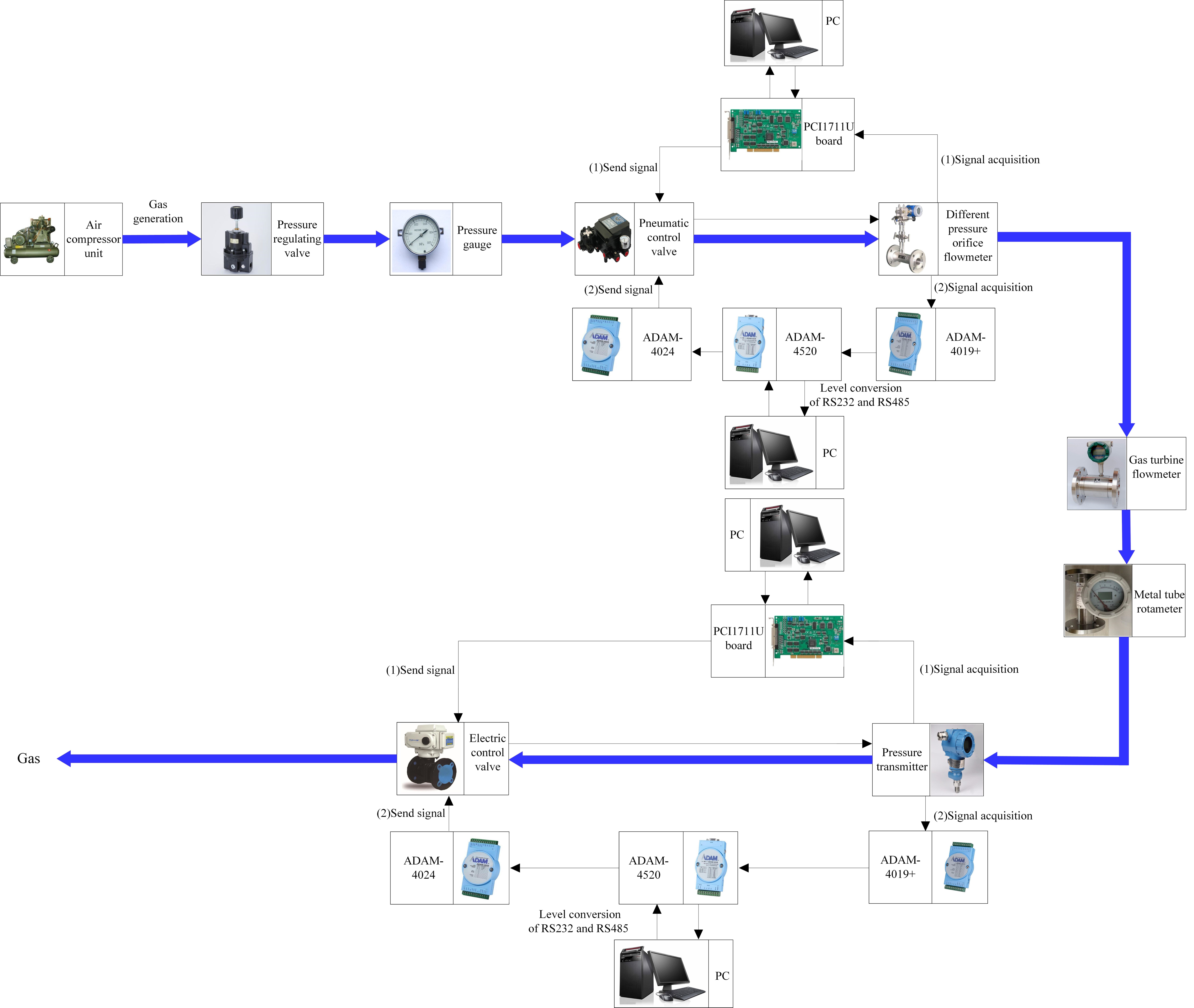}
\caption{The process principle of gas pipeline system}
\end{figure*}

\begin{figure}[htp]
\centering
\includegraphics[width=0.22\textwidth, height=0.22\textwidth]{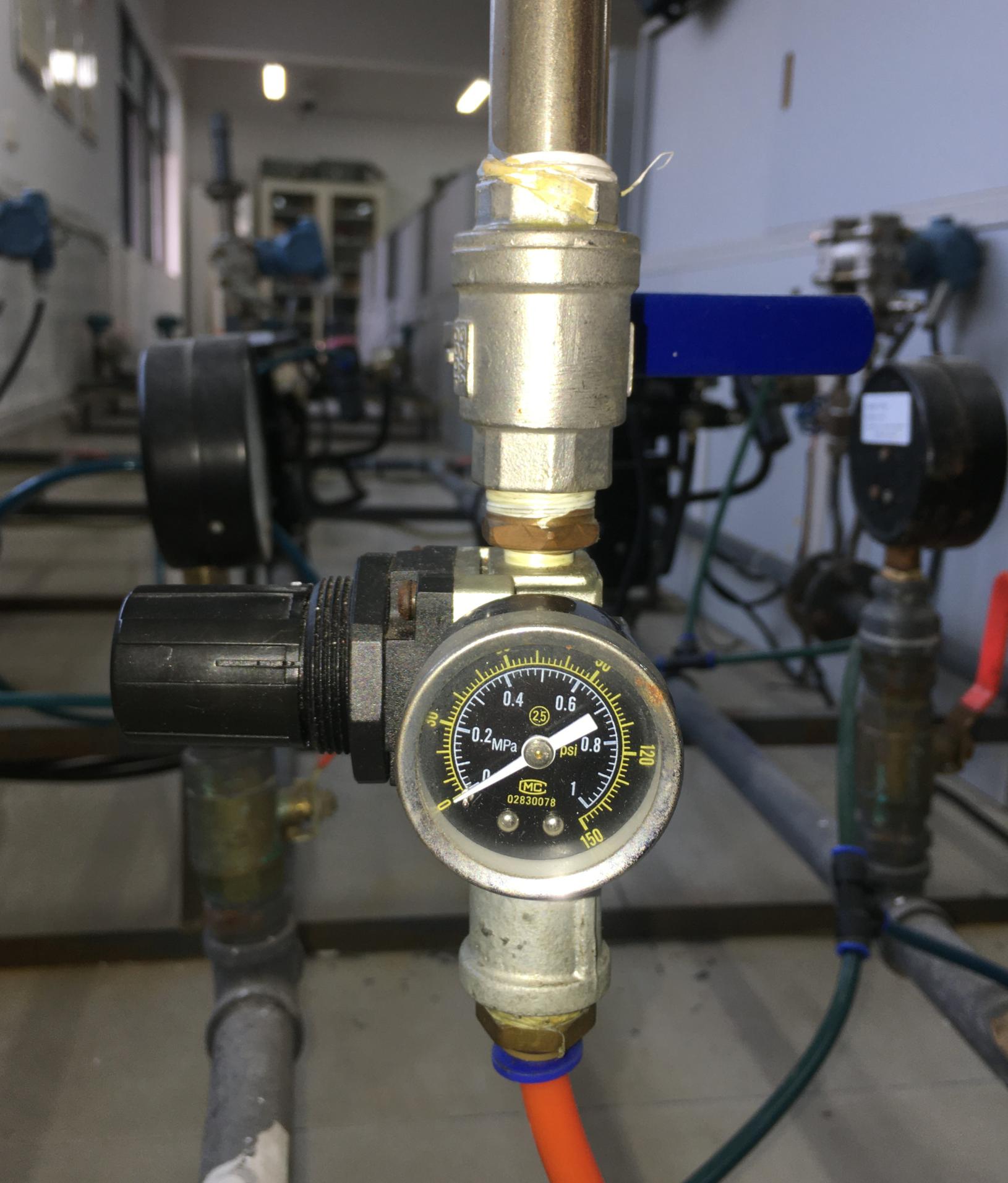}
\caption{Adjusting the air pressure of the pipeline}
\end{figure}

Finally, continuous and stable compressed air from air compressor units is transmitted to the front end of the gas pipeline system by the interconnecting piping, as shown in Fig. 1 (b). Considering that the back end of the gas pipeline system cannot be observed effectively in Fig. 1 (b), Fig. 1 (c) shows the details of the system backend. Moreover, in order to clearly describe the gas transmission process in Fig. 1, this paper gives a detailed process principle diagram of the gas pipeline system in Fig. 3. The gas pipeline system is equipped with pressure regulating valves, pneumatic actuators, electric actuators, sensors, signal acquisition devices, control devices, and so on. By adjusting the given value of pressure regulating valves in Fig. 4, the gas pressure in Figs. 1 (b) and 1 (c) can be changed synchronously. Then, leakage signals under different gas pressures (i.e., 0.2 Mpa, 0.4 Mpa, and 0.5 Mpa) can be collected through the self-developed device in Fig. 2.

\subsection{Object Categories}
\label{0.5page}

\begin{table*}
  \caption{Introduction of Each category in GPLA-12 dataset.}
  \label{Tab:bookRWCal}  \centering
\begin{tabular}{ccc}
  \toprule
   Label  &  File Name  & Description \\
\midrule
 1  &  $Data1\_0.2MP\_none$  & Gas pressure is 0.2 MPa in the pipeline with microphone 1 under noiseless environment  \\
     \midrule
2 &  $Data1\_0.2MP\_noisy$  & Gas pressure is 0.2 MPa in the pipeline with microphone 1 under strongly noisy environment \\
     \midrule
3 & $Data1\_0.4MP\_none$ & Gas pressure is 0.4 MPa in the pipeline with microphone 1 under noiseless environment \\
\midrule
 4  &  $Data1\_0.4MP\_noisy$  & Gas pressure is 0.4 MPa in the pipeline with microphone 1 under strongly noisy environment  \\
     \midrule
5 &  $Data1\_0.5MP\_none$  & Gas pressure is 0.5 MPa in the pipeline with microphone 1 under noiseless environment \\
     \midrule
6 & $Data1\_0.5MP\_noisy$ & Gas pressure is 0.5 MPa in the pipeline with microphone 1 under strongly noisy environment \\
\midrule
 7  &  $Data2\_0.2MP\_none$  & Gas pressure is 0.2 MPa in the pipeline with microphone 2 under noiseless environment  \\
     \midrule
8 &  $Data2\_0.2MP\_noisy$  & Gas pressure is 0.2 MPa in the pipeline with microphone 2 under strongly noisy environment \\
     \midrule
9 & $Data2\_0.4MP\_none$ & Gas pressure is 0.4 MPa in the pipeline with microphone 2 under noiseless environment \\
\midrule
 10  &  $Data2\_0.4MP\_noisy$  & Gas pressure is 0.4 MPa in the pipeline with microphone 2 under strongly noisy environment  \\
     \midrule
11 &  $Data2\_0.5MP\_none$  & Gas pressure is 0.5 MPa in the pipeline with microphone 2 under noiseless environment \\
     \midrule
12 & $Data2\_0.5MP\_noisy$ & Gas pressure is 0.5 MPa in the pipeline with microphone 2 under strongly noisy environment \\
    \bottomrule
  \end{tabular}
\end{table*}

GPLA-12 dataset is composed in two files (i.e., \textit{data.csv} and \textit{label.csv}), for which \textit{data.csv} stores the information of acoustic signals and \textit{label.csv} reserves the correponding labels. \textit{data.csv} has 684 rows and 1460 columns, and each row represents a sample. \textit{label.csv} contains 684 rows and 1 column.

The presented compilation is composed of labeled sets comprising 12 classes of acoustic gas leakage signals under various conditions. With the purpose of maintaing the balance among data types, classes included are arbitrarily chosen while the number and diversity of the dataset is limited. 

There are 12 classes in GPLA-12 dataset, and the details of these categories are shown in Table II. For convenience, these classes can be grouped in several loosely defined major parts:

\begin{enumerate}[]
\item 0.2 PM vs 0.4 PM vs 0.5 PM.
\item noisy vs noiseless.
\item microphone 1 vs microphone 2.
\end{enumerate}

The dataset involves a variety of acoustic signals under different conditions. This provides various options for different purposes (e.g., classification, prediction, and clustering) of both algorithms with shade and deep architectures, which is suitable for both binary and multiple classes problems.

\subsection{Data Description}
\label{1 page}
Each sample in our dataset can be denoted as a time-series sample. We choose 2 samples in different classes in the dataset randomly (as illustrated in Fig. 5). It seems that features of acoustic signals of gas pipeline leakage are not obvious in the time domain, which requires algorithms to have a good ability of feature extraction. The mapping from acoustic signals to fault type need to be discovered by the algorithms. 

\begin{figure}[htp]
\centering
\subfigure[]{\includegraphics[width=0.40\textwidth, height=0.13\textwidth]{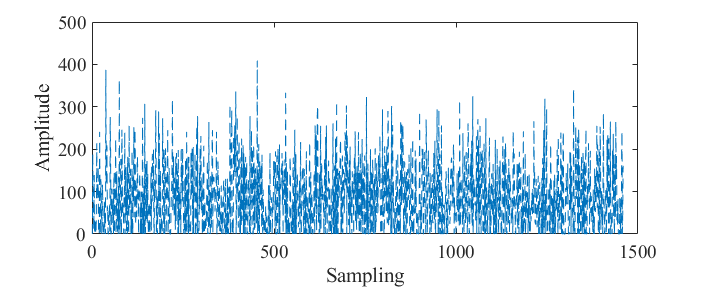}}
\subfigure[]{\includegraphics[width=0.40\textwidth, height=0.13\textwidth]{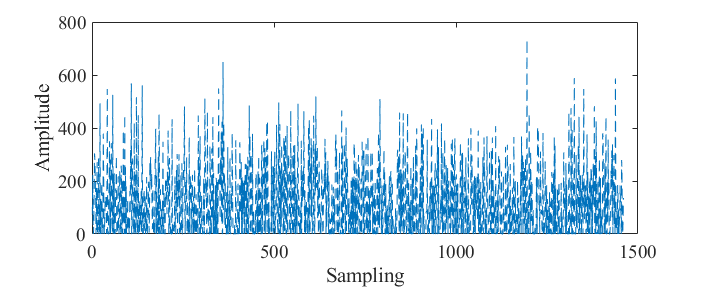}}
\caption{Samples in classes 2 and 4.}
\end{figure}

\section{Experiments}
\label{1 page}
In order to further study the dataset, we use our data to train both shallow classifiers in Scikit-toolbox \cite{scikit-learn} and deep structured classifiers: support vector machines (SVM), K-nearest neighbor (KNN), linear regression (LR), random forest (RF), AdaBoost, Gaussian Naive Bayes (GNB), Gaussian, Gradient Boosting tree (GBT), Rocchio, OneVsRest (OVR), recurrent neural network (RNN), convolutional neural network (CNN), and deep belief network (DBN). Additionally, we intend to testify the performance of baseline approaches and investigate the potential traps and intricacies. We use N-fold cross validation on the training/testing operations where N is 10 to obtain the performance. Meanwhile, we also divide GPLA-12 dataset randomly into training/testing data sets with a 7:3 ratio, and then train the classifiers. 

From Fig. 6, generally speaking, deep structured approaches operate effectively on extracting representative features. For acoustic signals of gas pipeline leakage, the deeply rich information are hard to be detected in the time domain. like most fault diagnosis cases (e.g., bearing \cite{xia2017fault, zhang2018deep, shao2017electric}, medical systems \cite{chen2019mechanical, wu2019intelligent}, chemical \cite{wu2018deep}, and gears \cite{jiang2018multiscale, jing2017convolutional}), CNN outperforms other methods on the basis of its learning ability of 1-D dimensional FDD data. Especially, GNB works well on GPLA datasets, originating from the Gaussian distribution characteristics of leakage acoustic signals while GNB has severe requirement of Gaussian distribution in the datasets. 

To explore details in the simulations, F1-score values of each class by different classifiers are illustrated in Fig. 7. Interestingly, for some classes, the F1-score are generally low, e.g., class 1. Moreover, some classifiers perform poorly for the dataset, such as KNN which displays badly on most classes leading to a bunch of 0. In other words, euclidean distance evaluation may not be suitable for GPLA-12 dataset.

\begin{figure}[htp]
\includegraphics[width=0.56\textwidth, height=0.2\textwidth]{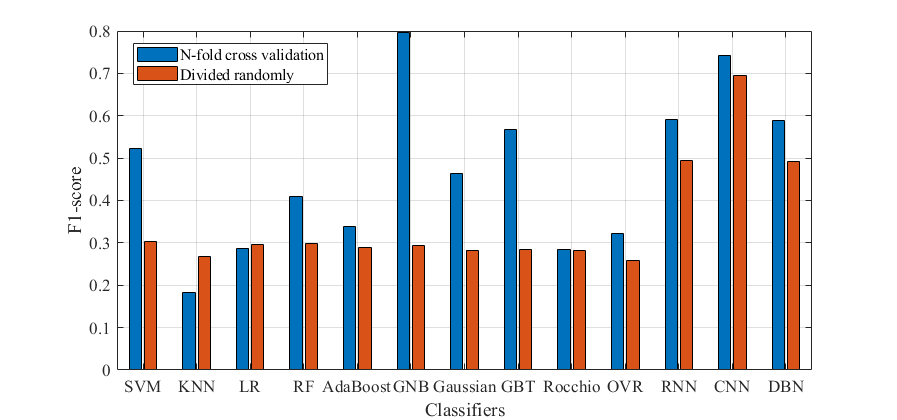}
\caption{Comparisons of F1-scores of classification with our dataset by different classifiers.}
\end{figure}

\begin{figure}[htp]
\centering
\includegraphics[width=0.33\textwidth, height=0.7\textwidth]{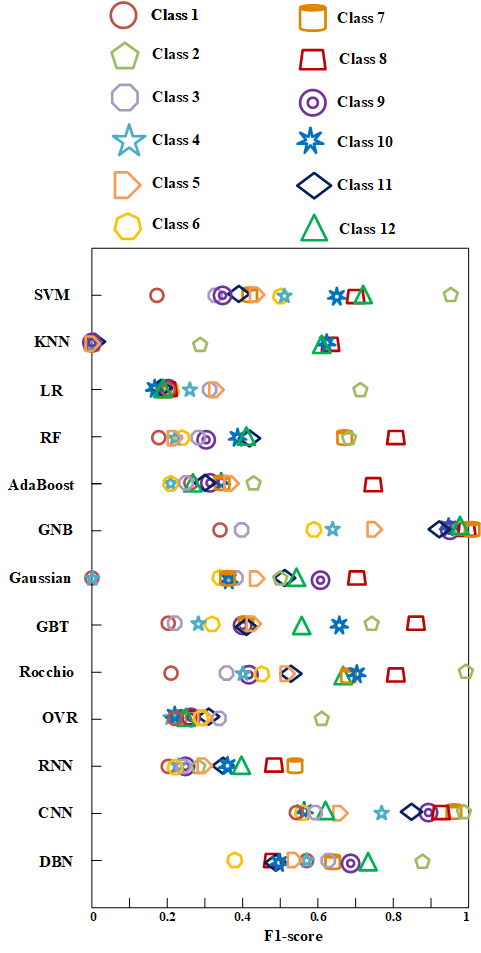}
\caption{F1-scores of each class by operating diverse classifiers on GPLA-12 dataset.}
\end{figure}

\section{Conclusions}
\label{+abstract 0.5}
The paper has provided a new dataset for fault diagnosis researches where publicly datasets are not abundant. The dataset can be used for binary and multi-class classification problems. In order to understand data feasibility, we have introduced the data collection system, the scale, and class information of the dataset. Moreover, we have discussed baseline classification results on this new dataset with both shadow and deep structured classifiers, where deep learning approaches are shown to have better performance due to their complex features of acoustic signals of gas pipeline leakage.

We have published the dataset and its information on the websites of www.daip.club and github.com/Deep-AI-Application-DAIP. We aim to update this dataset and expand both the quantity and type amount of the dataset regularly. Progress and results will be reported in due course.

\section*{Acknowledgment}
We would like to thank Min Gao and Hongcheng Liao for preprocessing the collected data, and Wenwen Zhu, Yifan Wang, and Keyu Chen for compiling the information online (i.e., www.daip.club and github.com/Deep-AI-Application-DAIP).


\bibliographystyle{IEEEtran}
\bibliography{IEEEabrv,file}

\end{document}